# Role of internal gases and creep of Ag in controlling the critical current density of Ag-sheathed $Bi_2Sr_2CaCu_2O_x$ wires

T. Shen[1, *], A. Ghosh[2], L. Cooley[1], J. Jiang[3]

1.  Technical Division, Fermi National Accelerator Lab, Batavia, IL, 60510
2.  Superconducting Magnet Division, Brookhaven National Lab, Brookhaven, NY, 11973
3.  Applied Superconductivity Center, National High Magnetic Field Laboratory, Florida State University, Tallahassee, FL, 32310

*Electronic mail: tshen@fnal.gov.

Abstract

High engineering critical current density $J_E$ of >500 A/mm$^2$ at 20 T and 4.2 K can be regularly achieved in Ag-sheathed multifilamentary $Bi_2Sr_2CaCu_2O_x$ (Bi-2212) round wire when the sample length is several centimeters. However, $J_E$(20 T) in Bi-2212 wires of several meters length, as well as longer pieces wound in coils, rarely exceeds 200 A/mm$^2$. Moreover, long-length wires often exhibit signs of Bi-2212 leakage after melt processing that are rarely found in short, open-end samples. We studied the length dependence of $J_E$ of state-of-the-art powder-in-tube (PIT) Bi-2212 wires and gases released by them during melt processing using mass spectroscopy, confirming that $J_E$ degradation with length is due to wire swelling produced by high internal gas pressures at elevated temperatures [1,2]. We further modeled the gas transport in Bi-2212 wires and examined the wire expansion at critical stages of the melt processing of as-drawn PIT wires and the wires that received a degassing treatment or a cold-densification treatment before melt processing. These investigations showed that internal gas pressure in long-length wires drives creep of the Ag sheath during the heat treatment, causing wire to expand, lowering the density of Bi-2212 filaments, and therefore degrading the wire $J_E$; the creep rupture of silver sheath naturally leads to the leakage of Bi-2212 liquid. Our work shows that proper control of such creep is the key to preventing Bi-2212 leakage and achieving high $J_E$ in long-length Bi-2212 conductors and coils.

# 1. Introduction

An important index of high-field superconducting magnet technology is the total current carried by a superconducting strand divided by its total cross-section area, or engineering critical current density $J_E$. High $J_E$ in magnetic fields of >20 T was demonstrated in $Bi_2Sr_2CaCu_2O_x$ (Bi-2212) multilayer, dip-coated tapes [3] and multifilamentary, powder-in-tube (PIT) tapes [4] processed by a melt processing approach [5], and then similar success was later reproduced in PIT multifilamentary Bi-2212 round wires [6-9]. Thus, melt-processed Bi-2212 wires are being considered for future very high field magnet technology, beyond the ~20 T limit of $Nb_3Sn$ magnet technology. As an example of the state of the art, $J_E$ reached 480 A/mm$^2$ at 4.2 K and 20 T in 2005 in a 1 m barrel sample made by Oxford Superconducting Technology [6,7], and this benchmark was recently pushed further to approximately 540 A/mm$^2$ at 4.2 K and 20 T in short pieces of ~10 cm wires [10]. A $J_E$ of 500 A/mm$^2$ is generally considered as sufficient for a new high-field magnet technology.

Despite these successes in obtaining high $J_E$ in short-pieces of Bi-2212, melt-processing of long-length Bi-2212 conductor often yields $J_E$ much less than expected [11-13]. $J_E$ of coils has rarely exceeded 200 A/mm$^2$ at 4.2 K and 20 T [11-13]. Moreover, melt-processed Bi-2212 coils have often suffered from leakage, showing clear discolorations where Bi-2212 liquid leaks through the encasing metal (Ag or Ag alloy) at high temperatures and reacts with surrounding materials such as insulation [11,14]. The leakage degrades $J_E$ as well as mechanical properties of superconductor composite [14-18]. These problems seriously limit magnet applications of Bi-2212. However, little has been known as regards to what causes leakage and $J_E$ degradation in long-length Bi-2212 and ways to eliminate the leakage and to improve the $J_E$ in coils have yet to be found because.

In the case of Bi-2212 tape conductor, one of major causes of low $J_E$ in long-length Bi-2212 was suggested to be high internal gas pressures at high temperatures [19-23]. Internal gases may include residual air and humidity present in as-drawn wires, gases released as a result of contaminants adsorbed on powder surfaces [24] and solid remnants in powder gasifying upon heating. Contaminants and gas impurities may include hydrates, hydroxides, carbonates [25], nitrates, and residues such as wire drawing lubricants (hydrocarbons); they transform to gases ($H_2O$, $CO_2$, or $NO_2$) upon heating. One evidence for this conjecture is that tapes with high carbon contents were observed to swell over long length, crack, or even bulge locally [23]. Gases can presumably diffuse along the conductor axis and out the ends when conductors are relatively short and are not sealed at the ends, but they would be trapped inside when conductors are relatively long or when their ends are sealed, creating high internal gas pressure and causing the conductor to deform irreversibly when heated at ambient pressure. The argument gives a plausible, qualitative explanation for $J_E$ degradation in long-length tapes, but details about contamination amounts, gas pressures, at what temperatures gases are being released are lacking.

For PIT round wires later fabricated from modern low-carbon powder (<200 ppm by weight, normalized to the weight of oxide filaments), the effect of trapped gas is more subtle and not yet clearly defined. For an Ag/AgCu sheathed multifilamentary Bi-2212 wire, Kuroda et al. [26] examined $J_E$ variation in short 3 cm open-end wire, long 1 to 3 m open-end wires, and a 10 m long coil. They found that: (1) $J_E$ in the 10 m long coil is 35% of that in the 3 cm, open-end wire; (2) $J_E$ in the 1.5 m open-end wire strongly varied along the length, with the middle section having the lowest $J_E$ and the end 20-30 cm section having $J_E$ comparable to that of the short, open-end 3 cm samples. Malagoli *et al.* [1] verified these findings in a more recently made Bi-2212 wire with a Ag-0.2wt.% Mg external sheath. Malagoli *et al.* [1] also made an important observation: the 1 m wire sample swelled along the entire length despite the ends being open during melt processing, and the greatest wire expansion occurred at the central section. Given the importance of removing porosity in PIT Bi-2212 for obtaining high $J_E$ revealed recently [10,27-32], the observation made by Malagoli et al. [1,2] naturally led to the conclusion that $J_E$ degradation in long-length Ag-sheathed wire is a direct consequence of the de-densification of Bi-2212 due to the wire swelling produced by the internal gas pressure.

While it has been tentatively established that internal gases are causing wire swelling in Ag sheath and $J_E$ degradation in long-length Bi-2212 wires, a number of important questions remain. The most crucial one concerns how silver deforms at high temperatures and at what temperatures the wire swells most significantly, since such understanding may shed light on what gases are primarily causing wire to expand or may even suggest whether possible remedies exist. Second, it is important to establish to what extent the low $J_E$ in long length wire is due to residual gas in green wires, and to what extent it is controlled by gases being released into wires during heat treatment, including $O_2$, $H_2O$, and $CO_2$, as such understanding will help assess whether an optimization in conductor design (such as using stronger sheath materials), or modification in wire manufacturing will reduce or eliminate the $J_E$ degradation. Third, it is important to establish a quantitative relationship between the rate of wire swelling and the gas impurity levels because the superconductor industry will need a clear idea of what levels of solid impurities can be tolerated for wires with different designs. Fourth, it is also important to have a quantitative description of gas transport in Bi-2212 wires so that the question as to what counts to be a long-length sample can be properly answered and the $J_E$ dependence on the conductor length could perhaps be predicted. We have addressed some of these issues by observing wire swelling for a group of representative state-of-the-art PIT multifilamentary round wires of Bi-2212 at critical stages of melt processing. Important and new information about the mechanism of wire swelling, leakage, and $J_E$ degradation has been obtained. We will also present a model of gas transport in Bi-2212 wires. The results of our studies are useful for developing Bi-2212 magnets that will consistently realize the full potential of this high field material for >22 T high field magnet applications.

## 2. Experimental design

The multifilamentary superconductor round wire selected for this study was manufactured by Oxford Superconducting Technology at New Jersey using the PIT technique by drawing a pure Ag tube with Bi-2212 precursor powder, then using a double restack to form several bundles. The final wire, 0.8 mm in diameter, contains hundreds of oxide filaments (37x18 filaments, ~20 μm in diameter) embedded in a matrix of pure silver, which is again encased in a stronger precipitation-hardening Ag-Mg sheath. The superconductor/Ag/AgMg ratios are 0.25/0.50/0.25. The filaments in as-drawn wires have an oxide packing density of ~74%. The wire was fabricated from a highly homogeneous, melt-cast precursor powder with a composition of $Bi_{2.17}Sr_{1.94}Ca_{0.90}Cu_{1.98}O_x$ developed by Nexans SuperConductors GmbH. Nexans indicated that the powder contained 30 ppmw carbon and 70 ppmw hydrogen. Such a 37x18, 0.8 mm wire has been demonstrated to carry a $J_E$ of 800 A/mm$^2$ at 4.2 K and 5 T (an extrapolation will gives a $J_E$ of >540 A/mm$^2$ at 20 T) [10] and has been investigated for making Rutherford cable for high-field magnets such as >16 T dipole.

Three key experiments were designed and performed on the same wire (batch PMM101108-2). First set of experiments, including two tests, were designed to quickly check whether internal gases are causing $J_E$ degradations. First we melt processed 8 cm and 1.2-1.4 m wires with their ends sealed and compared their $J_E$ and microstructure with those with their ends open during heat treatment. Sealing the ends of short-length wires before melt processing prevents gases from escaping and allows the behaviors of long-length wires to be studied. The heat treatment schedule for melt processing was to heat from room temperature to 820 °C at 160 °C/h, hold at 820 °C for 2 hours, heat again from 820 °C to 889 °C at 48 °C/h, hold at 889 °C for 0.2 hour, cool to 881 °C at 10 °C/h, further cool to 835 °C at 2.5 °C/h, hold at 835 °C for 48 hour, and then cool to room temperature. The processing was performed in 1 bar flowing O$_2$.

Critical current $I_c$ of fully reacted 8 cm wires were determined through the standard four point methods using an electric criterion of 1 μV/cm in a background field of up to 14 T. The $I_c$ variation along 1.2-1.4 m wire was examined using a method adapted from those established for testing critical current density of brittle Nb$_3$Sn conductors [33]. The 1.2-1.4 m long Bi-2212 strand was wound onto a grooved cylinder machined from a 96% alumina and then reacted. After the heat treatment, the spiral sample was transferred to a tubular barrel made of Ti-6Al-4V alloy and cooled down to 4.2 K for critical current measurement. Critical current along the length was determined by six pairs of voltage taps each monitoring current-voltage characteristics of 10 cm conductor at 4.2 K and fields up to 11 T at Brookhaven National Lab and to 14 T at Fermi National Accelerator Lab (Fermilab).

Second, gas release from Bi-2212 at high temperatures was monitored by analyzing the composition of gases in a vacuum tube furnace using a quadrupole residual gas analyzer while heating 5 cm Bi-2212 wires (with ends open) at 180 °C/h from 25 °C to 900 °C.

The second set of experiments we performed, including two tests, was central to this study. First, we observed the wire swelling and leakage patterns of wires through melt processing for 8 cm wires with ends sealed and compared to those with ends open. Second, we determined the rate of wire swelling at critical stages of melt processing for wires receiving various thermo-mechanical treatments. The wire swelling rate was measured by the rate at which wire diameters increased. Wire diameters were measured using a laser micrometer with a measurement resolution of 0.3 μm. The micrometer measures the diameters along eight axes by rotating the wire in steps of 22.5 °; an average wire diameter and a standard deviation are then recorded.

The third set of experiments was performed to gain more insights into the question of whether the wire expansion and leakage can be eliminated by reducing gases in the wires. We gave 8 cm and 1.4 m wires a degassing treatment at 450 °C in vacuum for 9 to 48 hours before melt processing (during vacuum annealing, sample ends were open to allow gases to escape.) The samples were then melt processed with their ends open or sealed, after which their diameter and $J_E$ along length were determined and analyzed, and compared to those samples that received no such treatments.

## 3. Results

### 3.1. Effects of closing wire ends

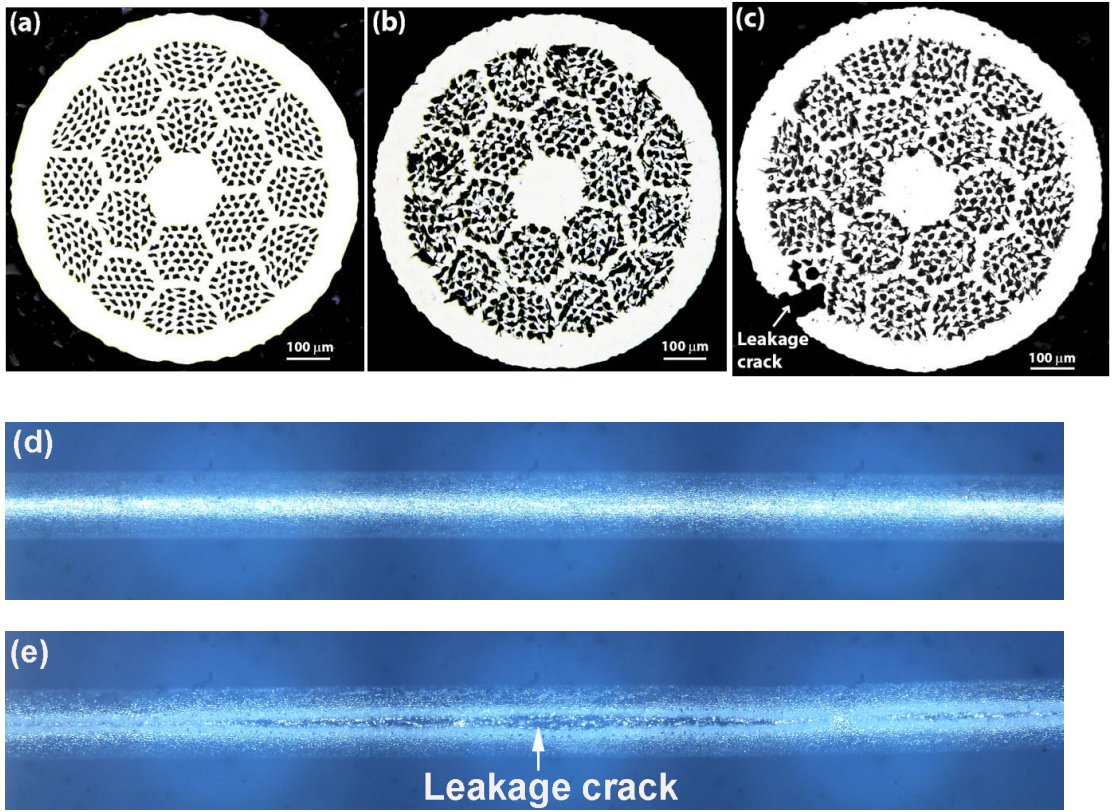

Figure 1: Representative micrographs of transverse cross-sections of (a) a unreacted Ag-sheathed multifilamentary Bi-2212 round wire, (b) an 8 cm long Bi-2212 wire after being melt processed with its ends open, (c) an 8 cm long Bi-2212 wire after being melt processed with its ends sealed. Photographs of sample (b) and (c) are shown in (d) and (e) respectively. Photograph (e) shows longitudinal cracks nearly parallel to the wire axis.

Figure 1 shows micrographs of a representative PIT Bi-2212 round wire before and after melt processing. Melt processing Bi-2212 wires with their ends sealed caused wire to swell: Diameter for sample (a) and (b) is 0.803 mm whereas the diameter for sample (c) increased to 0.830 mm. Such wire swelling is accompanied by significant critical current degradation: Wire (b) carried an $I_c$ of 180 A at 4.2 K and 5 T whereas $I_c$ of wire (c) scattered from 70 to 150 A. The closed-end sample leaked, showing longitudinal leakage cracks like the one in Figure 1.

Figure 2 shows the wire diameter as well as critical current $I_c$ along a 1.2-1.4 m wire. For open-end sample, wire swelling is the most pronounced in the middle section of the wire (by 1.5%) and decreased towards the ends (by 0.6%); the parabolic pattern of wire swelling was matched by a parabolic distribution of critical current density, which is the lowest in the middle section of the open-end wire. No leakage was found in 1.2-1.4 m open-end samples. Sealing the ends of the 1.2-1.4 m long wire caused it to swell more (by >2% in the center) and leak, and further reduced the wire $J_E$.

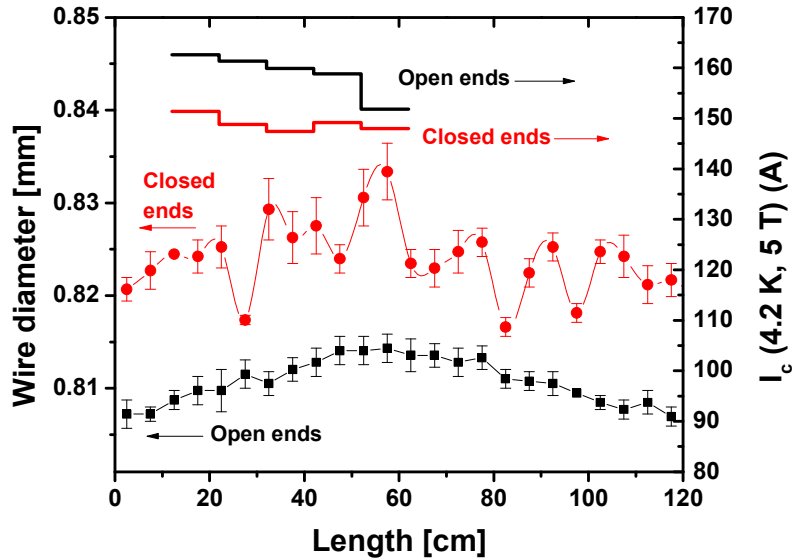

Figure 2: Variation of critical current $I_c$(4.2 K, 5 T) and wire diameter in 1.2-1.4 m long Bi-2212 wires melt processed with ends open or ends sealed. The wire diameter of sealed-ends sample show large error bars because wire cross-section became irregular as it leaked. The samples for diameter measurement were 1.2 m long whereas the samples for $I_c$ measurement were 1.4 m long. Only a half of the 1.4 m samples was measured for $I_c$ due to the expected symmetry.

### 3.2. Gas release by Ag-Bi-2212

Figure 3 shows the gas partial pressures in the vacuum furnace during heating 20 pieces of 5 cm long Bi-2212 wires at 180 °C/h to 900 °C. Major gases detected included $CO_2$, $CO/N_2$, $H_2O$, $H_2$, and $O_2$. A sudden increase of $O_2$ signal at 860-870 °C, an indication of the melting of Bi-2212 powder, was accompanied with an increase in $CO_2$ signal. On further increasing the temperature to 900 °C, partial pressures of $CO_2$, $H_2$, $H_2O$, and $N_2/CO$ fell whereas the partial pressure of $O_2$ persisted at high levels.

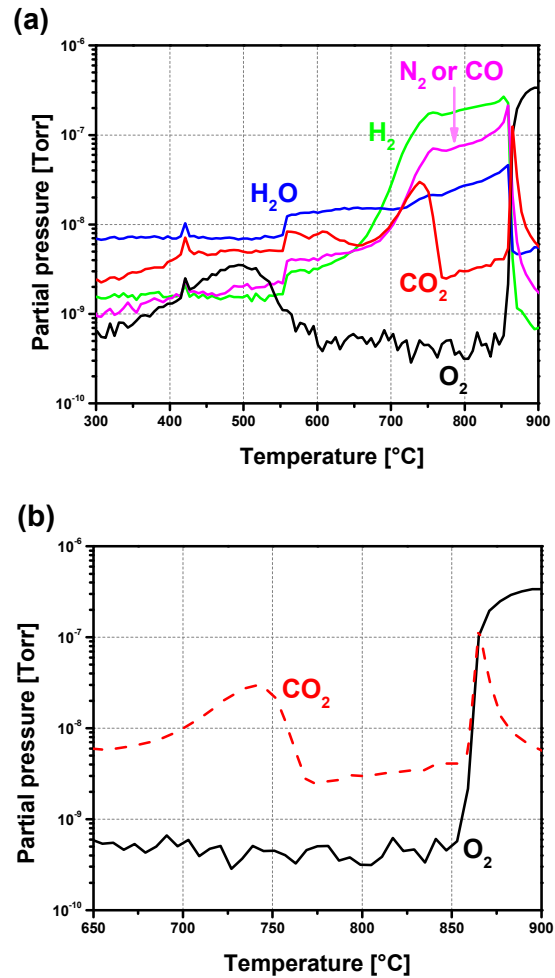

Figure 3: The gas partial pressure in a vacuum furnace detected by a residual gas analyzer. Inside the furnace were Bi-2212 round wires being heated at 180 °C/h to 900 °C. About 20 pieces of wire each 5 cm long with ends open were used for these measurements.

### 3.3. Rates of wire swelling and leakage at high-temperatures

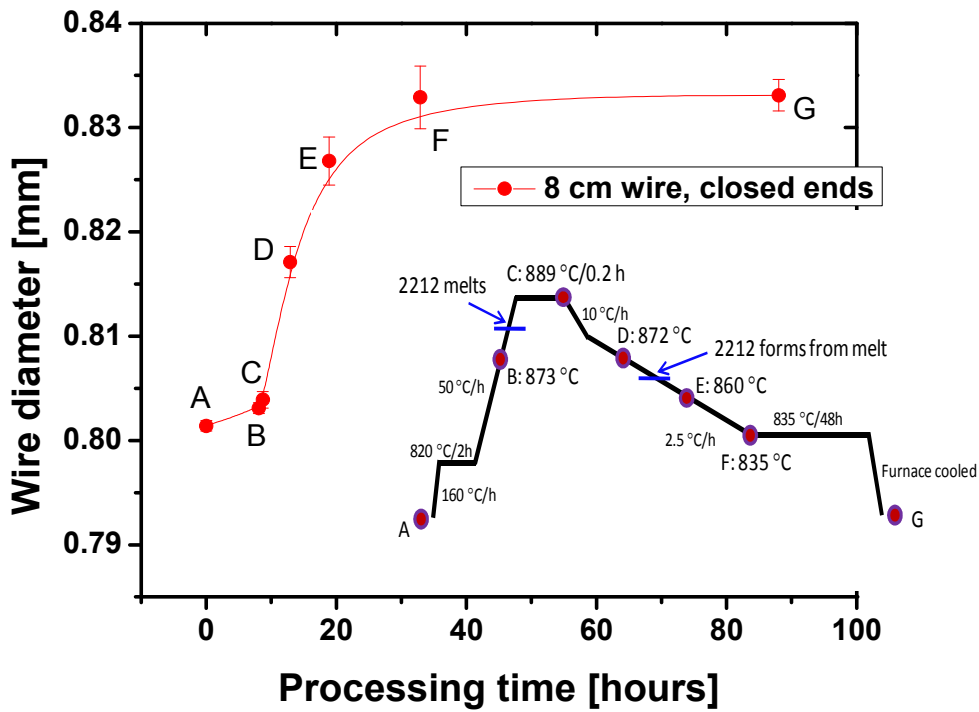

Figure 4: Progression of wire diameters of Bi-2212 along the melt processing. Wires were heat treated following the processing schedule to specific points and cooled down, after which their diameters were measured. The inset shows the melt processing used and critical processing points where the wire swelling was evaluated.

Figure 4 presents the progression of wire diameters of Bi-2212 along the melt processing. The diameter of closed-end samples increased over entire processing, whereas the wire diameter of open-end samples remained relatively unchanged. Wire swelling in closed-end samples accelerated after Bi-2212 melts around 880 °C on heating, peaked at 889 °C in the melt, decreased on cooling, and nearly stopped before 48 hour annealing at 835 °C. The leakage cracks were first observed in sample E.

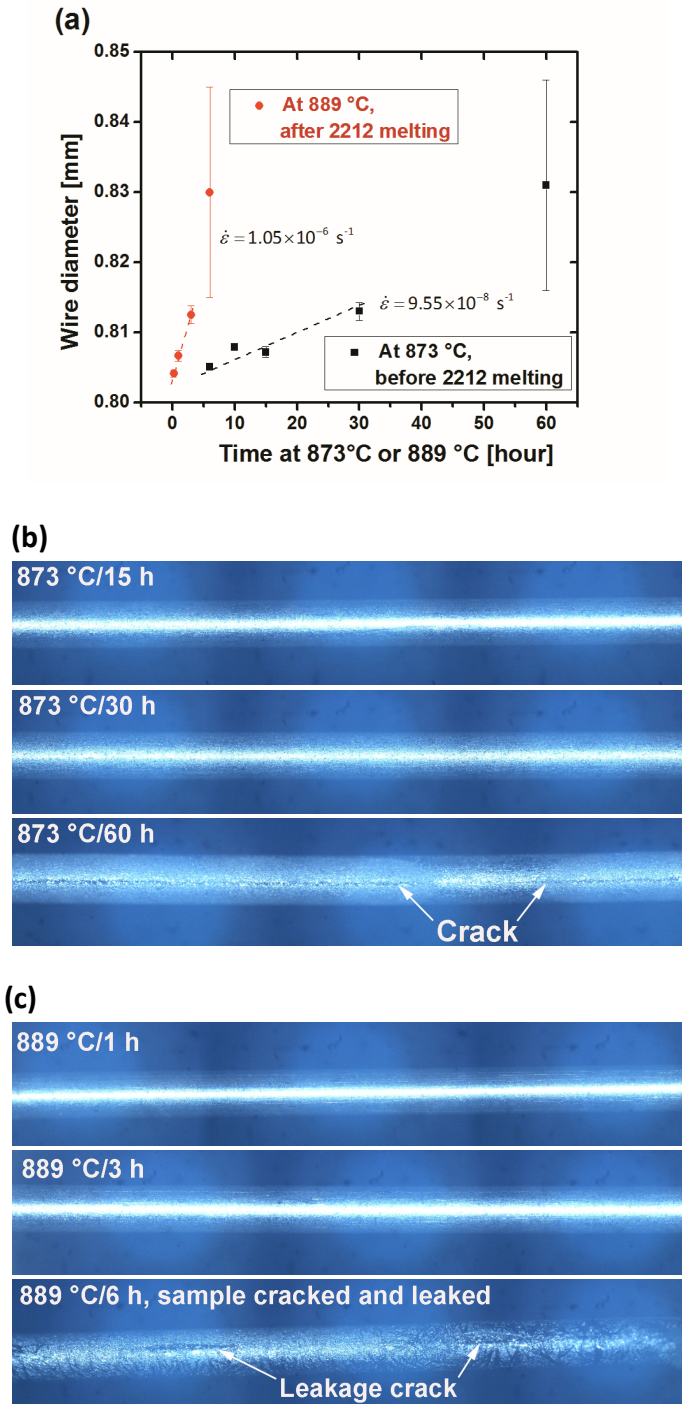

Figure 5: The rate of wire swelling at 873 °C (before melting), and 889 °C (after melting): (a) Wire swelling for 8 cm Bi-2212 sealed-ends wires at 873 °C and 889 °C.  Wires were heated from room temperature to 820 °C at 160 °C/h, annealed at 820 °C for 2 hours, and then heated at 50 °C/h to 873 °C or 889 °C, respectively.  Wires were then held either at 873 °C for up to 60 hours or at 889 °C for up to 6 hours and furnace cooled; sample diameters were then measured.  (b) Photographs of Bi-2212 sealed-ends samples cooled after being held at 873 °C for 15 hours, 30 hours, and 60 hours. (c) Photographs of Bi-2212 sealed-ends samples after being held at 889 °C for 1 hour, 3 hours, and 6 hours. In the case of (b) and (c), samples were photographed with the alumina paper on top of which samples laid during reaction.

Figure 5 presents the rates of wire swelling at two critical points of melting processing, at 873 °C before Bi-2212 melting and at 889 °C after Bi-2212 melting. Wire gradually swelled at both 873 °C and 889 °C. The wire diameter went up approximately at a rate of $9.55 \times 10^{-8}$ s$^{-1}$ at 873 °C but the rate increased by a factor of eleven to $1.05 \times 10^{-6}$ s$^{-1}$ at 889 °C. Wire held at 873 °C for 60 hours cracked (but didn't leak because Bi-2212 filaments have not yet been melted.) whereas at 889 °C wire cracked and leaked for only 6 hours. Wires were reacted on top of a 99.9% alumina paper. In the case of 889 °C – 6 hours, a clear reaction occur between leaked Bi-2212 liquid and alumina, whereas for 873 °C – 60 hours, wire cracks but there was no sign of such reactions.

### 3.4. Impact of degassing treatments

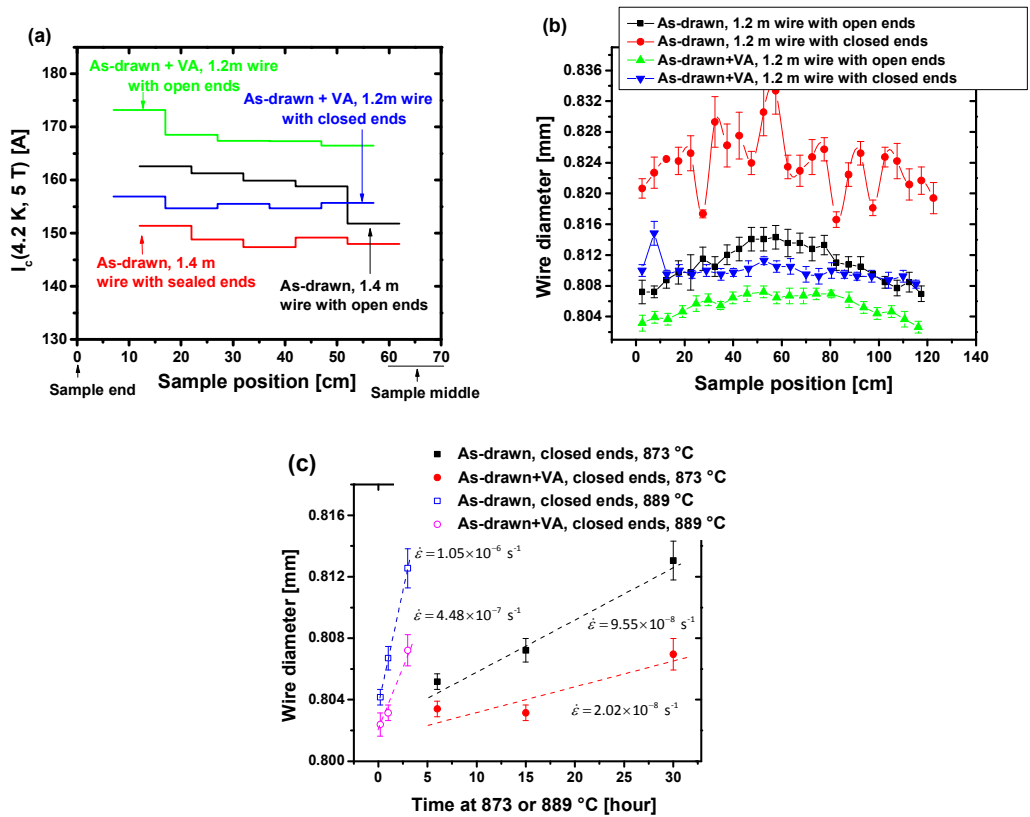

Figure 6: (a) The variation of $I_c$ in 1.2-1.4 m long wires that are melt processed after receiving a vacuum annealing at 450 °C for 9-48 hours, as compared to that of samples melt processed without a vacuum annealing. During vacuum annealing, the ends of all samples were open. (b) The wire diameter along the same group of samples. (c) The wire swelling at 873 °C or 889 °C for wires that received a vacuum annealing at 450 °C for 9 hours, as compared to that of samples that received no such treatments. VA=vacuum annealing.

Figure 6 shows that giving 1.2-1.4 m wires a vacuum annealing at 450 °C for 9 hours before melt processing raised wire $I_c$, while reducing the degree of wire swelling. Figure 6c also indicates that with vacuum annealing at 450 °C for 9 hours, the rate of wire swelling decreased to $2.02 \times 10^{-8}$ s$^{-1}$ at 873 °C and to $4.48 \times 10^{-7}$ s$^{-1}$ at 889 °C

# 4. Discussion

We determined $J_E$ and high-temperature expansion in a representative state-of-art Ag-sheathed PIT multifilamentary Bi-2212 round wire heat treated with ends open and ends closed, and confirmed the findings by Malagoli et al. [2] that $J_E$ of closed-end samples was severely degraded; in addition, the closed-end sample swelled, cracked, and leaked. The mass spectroscopy experiment indicated many gases, including $CO_2$ and $H_2O$, were being generated upon heating Bi-2212 wires. $J_E$ in 1.2 m open-end sample was not only degraded, but also strongly varied along the wire length, showing a parabolic pattern similar to that found by Kuroda *et al.* [12] and Malagoli *et al.* [1]. Applying a degassing treatment to Bi-2212 reduced the wire swelling and improved the $I_c/J_E$ uniformity in 1.2 m long sample; moreover, such a vacuum annealing treatment allows 1.2 m sample to reach >95% of the short-sample $I_c$. All of these findings are consistent with the underlying hypothesis that internal gases causes wire swelling and $J_E$ degradation [2].

Discussion of our experiments will proceed in three phases. First, we will associate the $J_E$ degradation of melt-processed long-length Bi-2212 wires to a decrease in the density of Bi-2212 filaments caused by wire swelling. Second, we will proceed in a more quantitative vein to understand how wire swelling occurs at high-temperatures under the influence of internal gas pressure by silver creeping and will conclude that the leakage is a natural consequence of the creep-rupture of Ag-alloy external sheath. Simple expressions will be given to describe the stress-sensitive nature of the creep-rupture behavior and time constants for gas diffusion in Bi-2212 wires. We will conclude with a discussion of the implications of our results for conductor development and high-field magnet engineering of Bi-2212.

## 4.1. $J_E$ degradation of long length wires caused by de-densification of Bi-2212 filaments

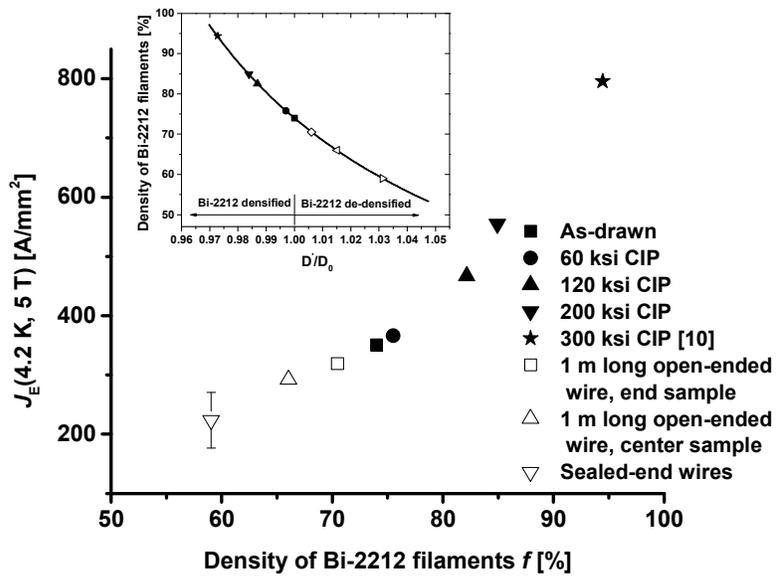

Figure 7: Changes in diameter and density of Bi-2212 filaments induced by cold densifications or wire swelling in a multifilamentary PIT round wire of Bi-2212 and their influence on its $J_E$(4.2 K, 5 T). $D_0$, the original diameter of the as-drawn wire, is 0.802 mm, whereas $D'$ is the new diameter of the wire after being subjected to various treatments. The relative density of Bi-2212 filaments $f$ ($f$=$V_{Bi-2212}$/( $V_{Bi-2212}$+$V_{porosity}$)) was calculated assuming incompressibility of Ag and AgMg and the initial packing density is 74% in as-drawn wires, other 26% of whose filament volumes are occupied by gases (porosity). CIP=cold isostatic processing, which was applied to as-drawn wires before melt processing. The as-drawn sample and CIP samples are 8-10 cm short samples melt processed with their open ends.

A direct consequence of the swelling of an Ag-Bi-2212 composite wire is that the density of Bi-2212 filaments will decrease. Figure 7 presents the calculated oxide filament density and $I_c$(4.2 K, 5 T) of representative long-length Bi-2212 samples that experienced various amount of de-densification, together with those of samples that received cold-isostatic pressing (CIP) powder compaction treatments. Figure 7 shows a strong proportionality between oxide filament density and $J_E$ in polycrystalline, melt processed Bi-2212 wires. A 1.2% decrease in the wire diameter under cold-isostatic pressing at 120 ksi increased the density of Bi-2212 filament from 74% to 82% , and as a result, its $I_c$ (open-end 8 cm sample) increased by 29%. For 1 m long open-end sample and closed-end samples, the oxide density decreased to be less than 59%, inducing degradation of $J_E$ by 55% (as compared to the as-drawn, open-end 8 cm sample). $J_E$ degradation in long-length wires is therefore due to the de-densification of Bi-2212 filaments.

The metallurgical interpretation of Figure 7 is that the key to extending high critical current density in short-length Bi-2212 samples to long-length wires is to eliminate the de-densification caused by wire swelling. The basis of porosity playing a major role in controlling connectivity and $J_c$ in Bi-2212 wires was given by the quench experiments combined with microscopy [27-31], which snapshot critical stages of melt processing and which revealed that upon melting, porosity in Bi-2212 filaments agglomerates into gas bubbles that are large and long enough to divide filaments into segments [28,29,31]. Though reshaped and redistributed at the latter stage of melt processing, these pores remain to block current flow in a fully processed wire. The importance of reducing porosity for achieving high $J_E$ found by quench experiments was further reinforced by Jiang et al. [10], who increased the Bi-2212 filament density of a multifilamentary PIT wire from 74% to 92% using CIP and doubled its $J_E$(4.2 K, 5 T) from 400 A/mm$^2$ to 800 A/mm$^2$. The accompanied quench studies and in-situ high-energy X-ray tomography studies of CIPed samples [32] revealed that CIPping significantly reduced the volume fraction as well as the size of gas bubbles in the melt.

## 4.2. The creep and rupture of silver sheath at high-temperatures as a result of Ag-Bi-2212 being subjected to high internal gas pressure

In this section, we will discuss how high internal gas pressures can lead to the swelling and rupture of silver sheath at high temperatures. Similar to a cylinder metal tube subjected to high internal pressures at high temperatures, a closed-end Ag/Bi-2212 wire would yield or creep under the influence of high internal gas pressure nearing the melting temperature of silver (920 °C in O$_2$) for a long duration time. Figure 5 shows that the Bi-2212 wires swell and then rupture, following a typical metal creep behavior that include three stages of plastic deformation: After a short primary creep stage, it went through a long period of steady-state creep, end with a negligible tertiary creep stage, and rupture. The temperature and stress dependence of the steady-state creep rate $d\varepsilon / dt = \dot{\varepsilon}$ can be described by a semi-empirical relation:

$$\dot{\varepsilon} = A \cdot \sigma^n \cdot e^{\frac{Q}{R \cdot T}}$$

Where $A$ is a constant, $n$ the stress exponent, $Q$ the activation energy, $R$ the gas constant, and $T$ the absolute temperature. The creep behaviors of pure Ag and dispersion strengthened AgMg under a simple tension or compression stress have been investigated by Leverant et al. [34], Nieh et al. [35], Goretta et al. [36,37]. Our discussion concerns the deformation of Ag and AgMg at >800 °C. The most relevant data was provided by Leverant et al. [34]. Their data show that the stress components are 4.34 for pure silver and 3.05 for Ag + 0.2 at. % MgO while the activation energies are 177.7 and 186.0 kJ/mol for pure silver and Ag + 0.2 at.% MgO respectively.

Here we summarize the steady-state creep rates determined for several representative samples in Table 1 (assuming that samples were in the steady-state creep state). Table 1 also indicated that the creep strain rate at 889 °C in the as-drawn wire is $1.05 \times 10^{-6}$ s$^{-1}$, which decreased to $4.48 \times 10^{-7}$ s$^{-1}$ in the wire that received a vacuum degassing treatment and which increased to $1.98 \times 10^{-6}$ s$^{-1}$ in the wire CIPped at 125 ksi before melt processing. Such changes in creep rates can be readily explained by changes in internal gas pressures. According to [34], an axial stress of 2.8 MPa is needed to produce a creep rate of $10^{-6}$ s$^{-1}$ at 889 °C under the simple tension test in pure Ag.

Table 1: Creep rates in wires that received different thermo-mechanical treatments at high-temperatures. Sample 1 is an 8 cm open-end, as-drawn wire whereas sample 2 is an 8 cm closed-end, as-drawn wire. Sample 3 is an 8 cm closed-end, as-drawn wire that was given a vacuum annealing at 450 °C for 9 hour before being melt processed. Sample 4 is 8 cm closed-end wire that was given a cold-isostatic pressing treatment at 125 ksi before being melt processed.

| Sample | Circumferential creep strain rates (s$^{-1}$) | | Leakage in fully processed wires? |
| --- | --- | --- | --- |
| | 873 °C, before Melting | 889 °C, in the melt | |
| 1 | Negligible | Negligible | No |
| 2 | $9.5 \times 10^{-8}$ | $1.05 \times 10^{-6}$ | Yes |
| 3 | $2.02 \times 10^{-8}$ | $4.48 \times 10^{-7}$ | No |
| 4 | $2.53 \times 10^{-7}$ | $1.98 \times 10^{-6}$ | Yes |

Long-time creep in metals and their alloys ends with inter-crystalline cracking and fracture substantially transverse to the direction of the principal stress. For our Bi-2212 round wire samples in which circumferential stress and radial stress are significant, long cracks parallel to the wire axis were observed (figure 1 and figure 5). We therefore conclude that the leakage of Bi-2212 is a natural consequence of the creep rupture of Ag sheath.

The creep-rupture strains were determined to be 12.57% for pure Ag and 1.31% for Ag + 0.1 at % Mg [35,37] at 400 °C and under an uni-axial stress of 35 MPa. This report showed that our wire had a rupture circumferential strain of 2.5-3.0%, indicated that this wire is limited by the ductility of AgMg. With the steady-state creep rate as large as $10^{-6}$ $s^{-1}$ for Bi-2212 wires at 889 °C, the time to rupture is small, about 5.6 hours. This prediction is quantitatively consistent with what was observed in figure 5, which showed wires leaked when held at 889 °C for 6 hours.

Many mathematical models have been developed to predict the stress state in metallic cylinder vessels subjected to internal pressures and/or external pressures at high temperatures and methods have also been developed to relate its circumferential strain rate to that given by a simple tension creep test. However, deducing the internal gas pressure from the creep rates in Table 1 is difficult because predicting the stress state and the creep rate in the multifilamentary round wire Bi-2212 is substantially more difficult due to the multifilament nature and the composite material structure.

### 4.3. Internal gases and their pressure in closed-end Bi-2212 wires at high-temperatures

Understanding that the fundamental cause of $J_E$ degradation and leakage in long-length Bi-2212 is the internal gas pressure, we may explore the nature of such internal gases. The first question to ask is to what extent the low $J_E$ in long-length wires is due to the residual gas in green wires. Our results implied that the residual gas unlikely plays a major role for two reasons. First, the residual gas analyzer clearly saw many gases were being released during heat treating as-drawn wires (figure 3). Second, if the residual gas is responsible for the Ag creep, the creep rate can't increase by a factor of 10 from 9.5 x $10^{-8}$ $s^{-1}$ to 1.05 x $10^{-6}$ $s^{-1}$ when increasing temperature from 873 °C to 889 °C. Such significant increase in the creep strain rate could be explained by the release of $CO_2$ during the melting of Bi-2212 as observed by Figure 3. $CO_2$ release at the melting of Bi-2212 powder may indicate that a major source of carbon gas is $SrCO_3$.

To roughly estimate the internal gas pressures in closed-end Bi-2212 PIT wires in the melt, we assume: (1) after Bi-2212 melts, 100% of C & H impurities are converted into gases; (2) the solubility of $CO_2$ and $H_2O$ in Bi-2212 liquid is negligible. As measurements indicated that W1 contained 134 ppmw hydrogen and 186 ppmw carbon, the internal gas pressure $P$ in the as-drawn PIT Bi-2212 wire according to ideal gas law (oxide packing density=74%; gas in unreacted wires is $O_2$) is 15.1 MPa, with $P(H_2O)$=12.2 MPa, $P(CO_2)$=2.8 MPa, and $P(O_2)$=0.1 MPa. Such high gas pressure is certainly strong enough to induce a creep rate in an order of $10^{-6}$ $s^{-1}$ in AgMg alloy, according to [34] [36,37]. This represents a ball-park analysis of the high-end pressure that wires could experience and we should note that at other processing regions, the internal gas pressure can be much lower. However, predicting such gas pressure can hardly be very accurate in the melt state or in other processing regions because of many complexities involved with melt processing a material like Bi-2212 during which gas phases may continue to react with other phases in the system.

### 4.4. Gas diffusion along wire axis in long-length Bi-2212 wires

For wires heat treated with their ends open, some fraction of gas can diffuse out through their ends. We will discuss how such gas diffusion leads to the $J_E$ degradation with increasing length and the strong variation of $J_E$ in a 1-2 m long wire as demonstrated by Figure 2 and Figure 6.

The gas transport in Ag-2212 composite conductor at high-temperatures is driven by gradients in gas molecule concentrations and amplified by gradients in internal gas pressure. We will first treat it as a gas diffusion problem without considering pressure effects, as for practical purposes, we do not need precise values of time constants of gas transport, and an appreciation of the orders of magnitude is sufficient.

The gas diffusion in a multifilamentary around wire of Bi-2212 along the filaments can be described by a one-dimensional diffusion equation:

$$\frac{\partial C}{\partial t} = \frac{\alpha^2}{L^2}\frac{\partial^2 C}{\partial x^2} \qquad 0 < x = \frac{x'}{L} < 1 \quad 0 < t < \infty$$

$$\text{Boundary conditions} \begin{cases} C(0,t) = 0 \\ C(1,t) = 0 \end{cases}$$

$$\text{Initial condition} \quad C(x,0) = C_0$$

where $C$ is the concentration of gas molecules, $\alpha^2 = D_e$ is the effective gas diffusion coefficient, $L$ is the length of the conductor being heat treated, $x'$ is the location coordinates, and x is dimensionless coordinates. The model here applies to the gases whose diffusivity through Ag matrix is negligible; the diffusion of oxygen through Ag will be separately discussed in next section.

Such partial-differential equation system can be solved analytically using the method of separation of variables. The solution is [38]:

$$C(x,t) = \sum_{0}^{\infty}\{A_{2n+1} \cdot e^{-[\frac{(2n+1)\cdot\pi\cdot\alpha}{L}]^2 \cdot t} \cdot \sin[(2n+1)\cdot\pi\cdot x]\}$$

$$\text{With } A_{2n+1} = \frac{4C_0}{(2n+1)\cdot\pi}$$

In long time, the terms further out in the series get small very fast due to the factor $e^{-[\frac{(2n+1)\cdot\pi\cdot\alpha}{L}]^2 \cdot t}$. Hence, for long time periods, the solution is approximately equal to the first term [38]:

$$C(x,t) \cong A_1 \cdot e^{-(\frac{\pi\alpha}{L})^2 \cdot t} \cdot \sin(\pi x)$$

$$\text{With } A_1 = \frac{4C_0}{\pi}$$

The time constant for gas diffusion is approximately as

$$t^* = \frac{L^2}{\pi^2 \cdot \alpha^2}$$

As we stated at the beginning, this diffusion process is accelerated by pressure gradients we have not yet taken into account so this $t^*$ is rather a conservative estimation for how quick gas transport can occur in Bi-2212.

Gas transport through a porous media like Bi-2212 powder mainly occurs by molecular diffusion and/or advection through the pores. Several empirical and semi-empirical perdition models can be used to estimate $D_e$. One of such models was proposed by Millington and Quirk (M-Q-model) [39,40]:

$$D_e = \theta_a \cdot T_a \cdot D_a^0$$

With $T_a = \frac{\theta_a^{7/3}}{n^2} = n^{1/3} \cdot (1 - S_r)^{7/3}$

Where $D_a^0$ is the free (undisturbed) diffusion coefficient in air ($D_a^0 = 1.8 x 10^{-5}$ m$^2$ / s for oxygen at 22 °C, $D_a^0 = 1.6 x 10^{-5}$ m$^2$ / s for CO$_2$ in air at 22 °C), $\theta_a$ is equivalent porosity, $T_a$ is the gas-phase tortuosity, $n$ is the porosity in the wire, and $S_r$ is the degree of saturation.

When 2212 is in powder state, $S_r$=0,

$$D_e = n^{4/3} \cdot D_a^0$$

In as-drawn wire, $n$=0.25. Using a $D_a^0 = 1.8 x 10^{-5}$ m$^2$ / s, the above equation gives $t^* = 9.9$ h for a 1 m long conductor at room temperature.

The M-Q model doesn't take into account gas diffusion through the liquid phase, so that for $S_r$=1 (saturated medium), it leads to an effective diffusion coefficient of zero ($D_e$=0), which is an unrealistic value but accurate enough for the case of gas diffusion through Bi-2212 liquid at high-temperatures, because a compound's diffusion coefficient is ~10$^4$ as great in air then in liquid.

$$t^*(\text{when Bi-2212 is in liquid state}) \cong 10^4 \cdot t^*(\text{when Bi-2212 is in powder state})$$

Approximating temperature dependence of diffusion coefficient as $D \propto T^{3/2}$ [Chapman–Enskog theory],

$$t^*(\text{in Bi-2212 liquid, 890°C}) \cong 13600 \text{ h} \quad \text{for 1 m conductor}$$

It is far greater than the time in the melt $t_{melt}$ for typical Bi-2212 melt processing, which is often less than 5 h so gas diffusion through liquid to two ends of wires can be omitted. This conclusion is verified by the residual gas analyzer data in the figure 3, which showed that the signals of CO/N$_2$, H$_2$, and CO$_2$ diminished after Bi-2212 melted.

Table 2 presented the estimated time constants $t^*$ for Bi-2212 wires in 0.1 m, 1.2 m, and 100 m.  One use of such data is to predict whether leakage will occur in a particular sample length.  Note that significant $CO_2$ released during the melting of Bi-2212 powder has only <10 minutes to diffuse out. Therefore, samples of >25 cm in length will trap some fractions of $CO_2$ so they are essentially no-longer a short-length sample.  Also, note that the melt processing we used allows samples to stay in the powder state above 800 °C for more than 3.5 hours, so in 1.2 m long or shorter open-end sample, the gas generated before 820 °C can diffuse out.  Further increasing wire length to >3 m will trap nearly all gases, producing a gas pressure comparable to that of the closed-end sample.  Therefore, leakage will begin to appear even in 3 m long open-ends sample.  All of the above predications on leakage were confirmed in representative PIT wires.

Table 2: Estimated time constants for gas diffusion
in Bi-2212 wires of 8 cm, 1 m, and 100 m in length

| Sample description | $t^*$ (hour) (800 °C) | Maximum $P_{total}$ Predicted (MPa) | Leakage occurrence |
|---|---|---|---|
| 8 cm, open-end sample | 0 | 0.4 | No |
| 8 cm, closed-end sample | - | 16.2 | Yes |
| 1.2 m, open-end sample | 2.12 | >2.90 but <16.2 | No |
| 100 m, open-end sample | 14700 | 16.2 | Yes |

## 4.5.  Oxygen diffusion through Ag matrix

It is worth discussing oxygen diffusion in Bi-2212 wires separately because it would diffuse through Ag quickly and also because there is big misconception in the literature (see for example [22]) that deformation of Ag matrix is dominantly caused by $O_2$ release.

When melted, Bi-2212 powder releases oxygen. From the thermal gravimetric data given by Kanai et al. [22] we can roughly estimated that the amount of oxygen released is >11800 ppmw.  Release of such large amount of oxygen into wire void space in a small time period of Bi-2212 melting has been argued to cause sudden increase in internal gas pressure, bulging Ag and causing "effluence of Bi-2212 liquid out of wire" [22]. Here we calculate the time constant for oxygen diffusion through Ag and argue that this is not the case.

The diffusivity and solubility of oxygen in solid silver follow the relations [41]:

$$D(\text{Oxygen in Ag}) = 4.9 \times 10^{-7} \cdot e^{\frac{11,600}{R \cdot T}} \ \text{m}^2/\text{s}$$

$$N(\text{Oxygen in Ag}) = 7.2 \cdot e^{-\frac{11,500}{R \cdot T}} \text{ at. percentage}$$

At 880 °C, D=1.5 x $10^{-7}$ m$^2$/s, N=2.17 at. %. The time constant $t^*$ for radial oxygen diffusion through Ag matrix an Ag-Bi-2212 composite wire in diameter of 0.8-1.2 mm at 880 °C is approximately 0.43-0.97 s ($t^*=L^2/(\pi^2*D)$, $L$=0.8-1.2 mm) s so O$_2$ can readily diffuse through Ag matrix, even when Bi-2212 is in melt state as seen in figure 3.

When processed in 100% O$_2$, Bi-2212 melting occurs predominantly in 880-886 °C. The melting event in our experiment (heating rate=0.8 °C/min) took $\approx$ 7.5 minutes, which should prevent pressure jump due to oxygen release.

### 4.6. Implications for conductor and magnet development and some final comments

Our data and analysis show that preventing creep of Ag driven by internal gases will prevent $J_E$ degradation and leakage in long-length Bi-2212 conductors. Further, our work reinforces the argument that high density in Bi-2212 filaments is essential for obtaining high wire $J_E$. Proper control of such creep to achieve high Bi-2212 density will allow development of long-length, leak-free, Bi-2212 wires that carry a $J_E$(4.2 K, 20 T) of >500 A/mm$^2$ for very high field superconducting magnet. Further conductor development should reduce or remove the source of internal gases through developing powders with lower gas impurity, adopting careful atmosphere control during wire processing, or using degassing treatments similar to that used in this study. Moreover, the creep rate may be decreased by increasing the thickness of the external AgMg sheath or replacing it with sheath materials with a higher creep resistance.

The above discussion and calculations help establish that the deformation behaviors of PIT Bi-2212 wires at high-temperatures can be described by creep of the silver sheath produced by high internal gas pressure, and help illustrate many aspects of the issue of $J_E$ degradation in long-length Bi-2212. Such analysis is useful for guiding Bi-2212 conductor development, but one has to bear in mind that the stress in the silver sheath is a complex function of wire architecture, gas impurities levels, heat treatment history, and porosity levels in Bi-2212 filaments when applying similar calculations to their wires for prediction of creep rates, leakage occurrence, and $J_E$ levels.

## 5. Conclusion

We have presented experiments and calculations that explicitly identified the internal gas pressure as the root cause of low $J_E$ as well as leakage in long-length Bi-2212 wire. We conclude that the $J_E$ degradation in long-length Ag-sheathed Bi-2212 wire is a direct consequence of de-densification of Bi-2212 filaments due to wire swelling produced by high internal gas pressures at elevated temperatures, confirming findings in [2]. We further demonstrated that the expansion of Ag-sheathed Bi-2212 PIT wires can be well understood in terms of silver creep at high temperatures. Leakage occurs by creep-rupture of the Ag-sheath and is the final manifestation of damages caused by internal gas pressure. Silver creep was due to strong hoop and radial stresses in silver matrix produced by high internal gas pressures.

This work, together with previous results [1, 2, 12], presents a conceptual framework to understand many puzzling experimental results such as the strong variation of $J_E$ in 1 m long sample and to predict if, when in the processing, and how a leakage would occur in a wire. The results of this study also have important consequences to the problem of attaining high critical current densities in long length Ag-sheathed multifilamentary conductors. Our data suggest that, by developing low gas impurity conductor and by controlling the silver creep to obtain high oxide filament density, it is feasible to develop long-length Bi-2212 wires that carries a $J_E$(4.2 K, 20 T) of >500 A/mm$^2$, which is 2.5 times higher than those being obtained presently in Bi-2212 coils; in addition, the conductor will be leak-free.

## Acknowledgement

Work at Fermilab was supported by the Office of Science at the U.S. Department of Energy (DOE) under contract No. DE-AC02-07CH11359. Work at Brookhaven National Lab was supported by the U.S. DOE under contract No. DE-AC02-98CH10886. T.S. was partially supported by a Fiscal Year 2012 U.S. DOE Early Career Research Program Award. We thank members of the Very High Field Superconducting Magnet Collaboration (VHFSMC) and the Bi-2212 Strand and Cable Collaboration (BSCCo), especially A. Tollestrup, D. Larbalestier, E. Hellstrom, Y. Huang H. Miao, J. Parrell, and S. Hong, for useful support and valuable discussions. We also thank M. Bossert, P. Li, E. Sperry and J. D'Ambra, D. Turrioni, J. Krambis, R. Mahoney, and M. Reynolds, for technical support.